# Picosecond photoresponse in van der Waals heterostructures

M. Massicotte[1], P. Schmidt[1], F. Vialla[1], K. G. Schädler[1], A. Reserbat-Plantey[1], K. Watanabe[2], T. Taniguchi[2], K.J. Tielrooij[1], F.H.L. Koppens[1]

[1] ICFO – Institut de Ciències Fotòniques, Mediterranean Technology Park, Castelldefels (Barcelona) 08860, Spain

[2] National Institute for Materials Science, 1-1 Namiki, Tsukuba 305-0044, Japan

**Two-dimensional (2D) crystals, such as graphene and transition metal dichalcogenides[1] (TMDs), present a collection of unique and complementary optoelectronic properties[2,3]. Assembling different 2D materials in vertical heterostructures[4] enables the combination of these properties in one device, thus creating multi-functional optoelectronic systems with superior performance. Here we demonstrate that graphene/$WSe_2$/graphene heterostructures ally the high photodetection efficiency of TMDs[5,6] with a picosecond photoresponse comparable to that of graphene[7–9], thereby optimizing both speed and efficiency in a single photodetector. We follow in time the extraction of photoexcited carriers in these devices using time-resolved photocurrent measurements and demonstrate a photoresponse time as short as 5.5 ps, which we tune by applying a bias and by varying the TMD layer thickness. Our study provides direct insight into the physical processes governing the detection speed and quantum efficiency of these van der Waals (vdW) heterostructures, such as out-of-plane carrier drift and recombination. The observation and understanding of ultrafast and efficient photodetection demonstrate the potential of hybrid TMD-based heterostructures as a platform for future optoelectronic devices.**

The optoelectronic response of 2D crystals is currently the subject of intense investigation[1–3,5–16] prompted by the need for next-generation photodetectors with superior performance in terms of efficiency, detection speed, as well as flexibility and transparency[17]. High photon absorption[5,18] and large photoconducting gain[11,12,14] have been observed in devices based on semiconducting 2D crystals. Yet, the observed response time typically ranges from nanoseconds[16] to seconds[11,14], with faster devices often displaying lower responsivity[11]. Therefore, the main challenge is to develop and assess photodetectors based on 2D semiconductor crystals that simultaneously possess a large active area, high internal efficiency, and fast response time.

A promising approach to create such versatile devices is to sandwich a TMD layer between two graphene sheets serving as charge extraction contacts. In contrast to lateral photodetectors, such vertical van der Waals (vdW) heterostructures[4] have the advantage of possessing a large, scalable active area and an atomically short charge extraction channel, potentially enabling both efficient and fast photodetection. Whereas the quantum efficiency of these vdW devices[5,6,13] and the dynamics of photocarrier creation and relaxation in TMDs[19-27] have been addressed, the response time of TMD-based photodetectors, as well as the dynamic processes governing their quantum efficiency remain elusive.

Here, we report on the intrinsic processes that limit the performance of photodetectors based on high-quality G/$WSe_2$/G (with G representing graphene) vdW heterostructures encapsulated in hexagonal boron nitride (hBN)[28]. We perform time-resolved photocurrent measurements[7,29] on devices consisting of $WSe_2$ flakes with a range of thicknesses (monolayer and multilayers from 2.2 to 40 nm). This technique, which combines electronic detection with subpicosecond optical excitation, allows probing of the extraction (Figure 1a) and loss dynamics of the photoexcited charge carriers in the photoactive TMD layer. We use $WSe_2$ as the photoactive material because of its high optical quality[15,30], and monolayer graphene flakes, placed below and above the TMD layer, as source and drain electrodes with gate-tunable work functions (Figure 1b). Figure 1c shows an optical image of a typical device comprising a WSe$_2$ flake with a thickness of 2.2 nm (3 atomic layers). Unless otherwise specified, all measurements presented in Figures 1 and 2 are obtained from this particular device.

We first characterize the photoresponse of our devices at room temperature using a scanning photocurrent microscope (see Methods). The photocurrent (PC) map displayed in Figure 1d shows that the photosensitive region corresponds to the area where the three main layers (G/WSe$_2$/G) are superimposed. The photocurrent spectrum (Figure 1e) measured in this area exhibits three main peaks (labelled *A*, *B* and *A'*) whose positions match well with the exciton absorption peaks of WSe$_2$[30–32]. This confirms that the photocurrent originates from the generation of photoexcited charge carriers upon light absorption in the WSe$_2$ layer, followed by the transfer of these charges to the graphene electrodes. The polarity and efficiency of the photocurrent are driven by the potential drop across the WSe$_2$ layer (*ΔV)*, which can be tuned by applying a bias ($V_B$) or gate ($V_G$) voltage[5,6] (see Supplementary Section I).

In order to investigate the dynamics of the photoexcited charges, we perform time-resolved photocurrent measurements[7,29] by exciting the photodetector with a pair of ultrashort pulses (~200 fs, centered at 1.55 eV) separated by a variable time delay *Δt* (Figure 1b, see Methods). Also known as photocurrent autocorrelation measurements - in analogy to autocorrelation measurements using non-linear optical crystals - this technique exploits the nonlinear power dependence of the photocurrent to extract the photocurrent dynamics. Figure 2a shows that this nonlinear power dependence is significant for average laser pulse intensities exceeding 10 kW/cm$^2$. The observed sublinearity likely originates from saturable absorption[22,33] (phase space filling), but we cannot exclude possible contributions from increased interactions between photocarriers (such as electron-hole recombination[29], exciton-exciton annihilation[21,25] and carrier-carrier scattering[22]), or from screening of the external bias (see Supplementary Section II). As a result of this sublinearity, photocurrent autocorrelation measurements, such as the one presented in Figure 2b, exhibit a symmetric dip around $\Delta t$ = 0. We ascribe the dynamics of the dip to depletion of photocarriers from the active region[29], which we can describe (see Methods) with a characteristic time constant $\tau$ that corresponds to the photoresponse time of our WSe$_2$-based photodetector. We will show that regardless of the mechanism that leads to the sublinear power dependence, the extracted time constant is representative for the photophysics that takes place in the device.

We determine $\tau$ for five devices with different WSe$_2$ layer thicknesses $L$ (Figure 3a) and find that $\tau$ increases with $L$ and varies by more than two orders of magnitude: from 10 ns (40 nm device) down to 5.5 ps (monolayer and 2.2 nm devices). Figure 3b furthermore shows that $\tau$ scales quadratically with the thickness $L$ (except the monolayer, which we discuss later). The highest response rate of $\Gamma = \tau^{-1}$ = (5.5 ± 0.1 ps)$^{-1}$, corresponding to a bandwidth of $f = 0.55/\tau$ = 100 GHz, is comparable to the intrinsic photo-switching speed of typical graphene detectors (~260 GHz)[7]. We note that in general ultrafast devices may have an operating speed limited by their resistance-capacitance (RC) time. In our devices the relevant capacitance is that of the WSe$_2$ channel and the resistance is the sum of the graphene resistance and graphene-metal contact resistance. We indeed expect the photoresponse time to be RC-limited to $\tau$ > 0.4 ns for the thinnest devices (L < 10 nm, see Supplementary Section III). Using real-time electronic measurements, we find an instrument-limited higher bound of ~1.6 ns. This is already significantly faster than any TMD-based photodetectors reported to date[6,11,12,14] and can further be improved by optimizing the circuit design, graphene resistance and contact resistance, up to the intrinsic limits that are reported in this work.

To address the underlying physics governing the dynamics and efficiencies in more detail, we study the dependence of $\tau$ on the bias voltage $V_B$ (Figure 3c) and observe that $\tau$ decreases with $V_B$ (Figure 3c, inset), indicating that photocarriers escape the active region more quickly as the electric field across the WSe$_2$ layer increases. For devices made of a relatively thick WSe$_2$ layer ($L \geq 7.4$ nm), the response rate $\Gamma$ depends linearly on $V_B$ (Figure 3d). This observation and the scaling of $\tau$ with $L^2$ suggests out-of-plane drift-diffusive transport: to reach the graphene contact, photocarriers generated in the middle of the WSe$_2$ layer travel a distance $L/2$ with a drift velocity $v_d = \mu E$ where $\mu$ is the out-of-plane photocarrier mobility and $E$ is the electric field across the WSe$_2$ layer. Assuming that $E \approx V_B/L$ (see Supplementary Section IV), the time $\tau_d$ it takes for photocarriers to drift out of the active region is $\tau_d = L^2/2\mu V$. This simple expression, represented by the dotted line in Figure 4a, captures the trend of $\Gamma$ at low $V_B/L^2$ (<$10^{-2}$ V/nm$^2$), yielding $\mu = 0.010 \pm 0.003$ cm$^2$/Vs. This value is similar to the one recently reported for thick MoS$_2$ flakes obtained from transport measurements[34], and is consistent with the strong conductivity anisotropy observed in TMDs (the in-plane conductivity is typically 2-3 orders of magnitude larger than the out-of-plane conductivity)[35].

Interestingly, for higher values of $V_B/L^2$ (>0.1 V/nm$^2$) the response rate reaches a limit. For $L = 2.2$ nm (trilayer), this occurs for high $V_B$ ($V_B > 0.6$ V), whereas the monolayer device reaches this limit without any applied bias. This saturation behaviour indicates the existence of an additional process that occurs in series with the drift process, thus prolonging the carrier extraction process. We find that the timescale $\tau_s$ of this additional process is bias-independent and is about 3-5 ps. This value is similar to the transfer time of photocarriers at the interface between WSe$_2$ and graphene (~1 ps)[20] and to the hot exciton dissociation time in few-layer MoS$_2$ (~0.7 ps)[27], which were both recently measured in all-optical pump-probe experiments. We thus argue that the response rate at high $V_B/L^2$ has a higher bound that is determined by these two processes rather than by drift (Figure 4b); this constitutes an intrinsic limitation for this device geometry.

We also observe a lower bound for the response rate, occurring at low $V_B$, which we attribute to a loss mechanism that occurs in parallel to the photocurrent generation process (Figure 4b). We find that this process presents a strong thickness dependence, with time constant $\tau_r$ ranging from 40 ps (2.2 nm device) up to >10 ns (40 nm device). We attribute this effect to radiative and non-radiative recombination of photocarriers[19,23,24,26] and energy transfer via dipole-dipole interaction to the graphene sheets[36] - two processes that have been shown to depend on TMD layer thickness. The latter interpretation is corroborated by the quenching of the WSe$_2$ photoluminescence observed in the trilayer (10-fold quenching) and monolayer (300-fold quenching) devices (see Supplementary Section V).

We now show that the internal quantum efficiency (IQE, defined as the ratio between the number of extracted photocarriers and the number of absorbed photons) of the photodetector is the direct result of the competition between the photocarrier extraction time ($\tau_d + \tau_s$) and recombination loss ($\tau_r$). Figure 4c shows the IQE as a function of $\Delta V$ (the potential drop across the WSe$_2$ layer) that can be controlled by both $V_G$ and $V_B$ (inset in Figure 4c). We find that the experimental IQE matches well the extraction efficiency $\tau_r/(\tau_d + \tau_s + \tau_r)$ derived from the dynamic model illustrated in Figure 4b. This concordance confirms that the dynamic processes identified in our time-resolved study correspond to the relevant physical mechanisms that govern the photoresponse of the device in the linear response regime (low-power, quasi-continuous excitation). It further demonstrates that the high IQEs observed in TMD-based vertical photodetectors (>85%)[6] arise from the short extraction time of the photocarriers out of the thin TMD channel, thereby outcompeting the loss mechanisms.

This suggests that the IQE can be optimized by reducing the channel length, and thus minimizing the TMD thickness. Interestingly, for the most extreme case of a WSe$_2$ monolayer, we observe a very low IQE (~6%, assuming 5% absorption) despite of a very short response time. This indicates that the observed dynamics in this device, as opposed to thicker ones, correspond to the losses, rather than to the extraction of charges. This is supported by the observation that the response time is bias-independent. The two loss mechanisms mentioned previously (intrinsic recombination and energy transfer) should indeed both occur much faster for a monolayer: recombination is facilitated by the direct nature of the bandgap of monolayer WSe$_2$[19,30], and loss through energy transfer exhibits a strong power law dependence with the inverse dipole-dipole distance[36].

We conclude that G/WSe$_2$/G devices made of trilayer WSe$_2$ (2.2 nm) offer the best compromise for optimizing the IQE as they exhibit both a fast photoresponse (down to 5.5 ps) and a high internal quantum efficiency (>70%). Moreover, the device can be efficiently operated without dark current, and thus with low noise level, by applying a large gate voltage and no bias voltage (Noise-equivalent Power down to ~5x10$^{-12}$ W/Hz$^{1/2}$, see Supplementary Section VI). Further improvement of the external quantum efficiency (EQE = 7.3% in our 2.2-nm-thick device) can be achieved by enhancing the light-matter interaction in the TMD layer with, for instance, optical waveguides[9,11] and cavities or plasmonic nanostructures[5,10]. Stacking multiple semiconducting 2D crystals with different bandgaps is also a promising way to extend the spectral absorption range[2] without compromising much the photoresponse time. Our comprehensive understanding of the photocurrent dynamics in G/TMD/G heterostructures paves the way for designing vdW heterostructures for fast and efficient optoelectronic applications, such as high-speed integrated communication systems.

## Acknowledgements:

The authors are grateful to Qiong Ma and Pablo Jarillo-Herrero for their instruction on the layer assembly technique, and Mark Lundeberg for valuable discussions. MM thanks the Natural Sciences and Engineering Research Council of Canada (PGSD3-426325-2012). FV acknowledges financial support from Marie-Curie International Fellowship COFUND and ICFOnest program. F.K. acknowledges support by the Fundacio Cellex Barcelona, an ERC Career integration grant (294056,GRANOP) and ERC starting grant (307806, CarbonLight) and support by the EC under the Graphene Flagship (contract no. CNECT-ICT-604391).

## Author contributions:

M.M. and F.H.L.K. conceived and designed the experiments. M.M., P.S. and F.V. fabricated the samples, carried out the experiments and performed the data analysis. K.W. and T.T provided boron nitride crystals. K.G.S. and A.R.P. provided assistance for the photoluminescence measurements. M.M., F.V., K.J.T., P.S. and F.H.L.K co-wrote the manuscript, with the participation of K.G.S. and A.R.P.

## Additional information:

Supplementary information accompanies this paper at www.nature.com/naturenanotechnology. Reprints and permission information is available online at http://npg.nature.com/reprintsandpermissions/.
Correspondence and requests for materials should be addressed to F.H.L.K.

## Competing financial interests:

The authors declare no competing financial interests.

**Figure captions**

*Figure 1* **Photocurrent generation in G/WSe$_2$/G heterostructure. a)** Schematic representation of photoexcited charge carrier dynamics in a hBN/G/WSe$_2$/G/hBN heterostructure. Graphene and hBN layers are respectively coloured in black and green, W and Se atoms in blue and orange. Upon pulsed-laser excitation, excitons or electron-hole pairs are created, separated and transported to the graphene electrodes. **b)** Schematic illustrating the experimental time-resolved photocurrent setup and cross-sectional view of the device. Two ultrashort pulses, delayed by a controllable distance *cΔt* (*c* denotes the speed of light), are focused on the device, which comprises a backgate ($V_G$). A bias voltage ($V_B$) can be applied between the top ($G_T$) and bottom ($G_B$) graphene layers through which photocurrent is measured (see Methods). **c)** Optical image of a heterostructure comprising a 2.2 nm thick WSe$_2$ flake. The graphene and WSe$_2$ flakes are outlined and shaded for clarity. **d)** Photocurrent (PC) map obtained by scanning a focused laser beam with a wavelength *λ* = 759 nm and a power *P* = 5 µW on the device shown in **c** with $V_G$ = 0 V and $V_B$ = 0.4 V. The photocurrent is mainly generated in the region where graphene layers (delimited by grey dotted lines) overlap. **e)** Responsivity spectrum of the device shown in **c** measured at a constant power *P* = 5 µW, with $V_G$ = 0 V and $V_B$ = 0.4 V. The exciton peaks are labelled according to the convention of Wilson and Yoffe[31] for WSe$_2$ and confirm that the PC stems from the vertical extraction of carriers generated in the WSe$_2$ layer. The orange shading indicates the spectral range of the pulsed excitation (illustrated in **b**) employed for time-resolved measurements.

*Figure 2* **Extraction of the photoresponse time of a G/2.2-nm WSe$_2$/G heterostructure by time-resolved photocurrent measurements. a)** Photocurrent (*PC*) vs. laser pulse power (*P*) for $V_B$ = 0.2 V. The transition from the linear to sublinear regime takes place at *P* ∼ 100 µW (∼10 kW/cm$^2$). The dotted line corresponds to a linear relationship between *PC* and *P*. Inset: the same data presented on a linear scale. **b)** Photocurrent as a function of time delay between two pulses (illustrated above the plot) with *P* = 300 µW, at $V_B$ = 1.2 V (blue dots). The photocurrent is normalized by the value of the photocurrent saturating at long *Δt*. The solid black line is a fit to the data using the model described in Methods, yielding a time constant *τ* = 5.5 ± 0.1 ps. Inset: Schematics of the photoresponse time in a G/WSe$_2$/G heterostructure, here represented by a band diagram with a bias voltage $V_B$ applied between the graphene layers. The red sinusoidal arrows, blue dot and blue circle symbolize photons, photoexcited electron and hole, respectively. The photoresponse time *τ*, which corresponds to the time that photocarriers reside in the photoactive area, is represented by the black arrows.

*Figure 3* **Tuning of the photoresponse time $\tau$ by variation of the WSe$_2$ layer thickness $L$ and bias voltage $V_B$. a)** Time-resolved photocurrent measurements on heterostructures with different $L$ (curves are offset for clarity). All measurements are obtained using a laser pulse power $P \sim 300$ µW and a bias voltage $V_B = 0.5$ V, except the monolayer which was measured at $V_B = 0$ V. Solid black lines are fits to the data. **b)** Photoresponse rate $\Gamma = 1/\tau$ vs. $L$. Data points represent the averaged values of $\Gamma$ for different $V_B$ (typically from 0.1 to 1 V), whereas the error bars correspond to the minimum and maximum value of $\Gamma$. The solid black line is a power law fit (excluding the monolayer device), yielding $\Gamma \propto L^{-1.9 \pm 0.3}$. **c)** Time-resolved photocurrent measurements on a heterostructure with $L = 7.4$ nm at various bias voltages $V_B$ and $P = 300$ µW (curves are offset for clarity). Inset: photoresponse time $\tau$ extracted from this heterostructure as a function of $V_B$. **d)** Photoresponse rate $\Gamma$ as a function of $V_B$ for various values of $L$. The solid black line corresponds to a linear relationship between $\Gamma$ and $V_B$, whereas the dotted line labels the effective minimum response time $\tau = 5.5$ ps.

*Figure 4* **Dynamic processes governing the photoresponse of G/WSe$_2$/G heterostructures. a)** Response rate $\Gamma = 1/\tau$ vs. $V_B/L^2$ (same data as Figure 3d). The dotted line corresponds to the diffusive transport model discussed in the main text with photocarrier mobility $\mu = 0.010 \pm 0.003$ cm$^2$/Vs. The uncertainty in $\mu$ is estimated from the small variations in the mobility of each device, which, among other things, could stem from extrinsic factors (e.g. unintentional doping) or variations in the laser power employed to measure each device (see Supplementary Section VII). The solid black lines correspond to the equation $\tau^{-1} = (\tau_d + \tau_s)^{-1} + \tau_r^{-1}$ with $\mu = 0.01$ cm$^2$/Vs, $\tau_s = 3$ ps and the values of $\tau_r$ indicated for each line. **b)** Schematic of the processes contributing to the response time. Radiative and non-radiative photocarrier recombination ($\tau_r$) leads to losses, whereas photocurrent is generated by photocarrier drift ($\tau_d$) in combination with exciton dissociation and/or charge transfer at the G/WSe$_2$ interface ($\tau_s$). **c)** IQE as a function of the potential drop across the WSe$_2$ layer $\Delta V$ for devices with L = 2.2, 7.4 and 28 nm. The black lines correspond to the expression $\tau_r/(\tau_d + \tau_s + \tau_r)$, where $\tau_d = L^2/2\mu\Delta V$, using the values indicated in **a.** The experimental IQE (data points) is obtained by measuring the photocurrent generated at low power ($P = 1$ µW) with a quasi-continuous laser ($\lambda = 759$ nm) and by using the absorption as a fit parameter. The fitted absorption values are shown in the top left inset and compare well with the absorption expected from Beer-Lambert's law (solid black line) using the bulk absorption coefficient of WSe$_2$ ($3 \times 10^5$ cm$^{-1}$)[32]. Bottom right inset: IQE vs $V_B$ measured on the 7.4-nm-thick device under the same illumination conditions as the main figure, with $V_G$ from 0 to 60 V in steps of 20 V.

## Methods:

**Device fabrication.** Our vdW heterostructures are prepared using a layer assembly technique similar to that reported by Wang *et al.*[28]. First, hBN is exfoliated onto a polymer (PMMA or PPC) and one of the flakes is employed to successively pick up graphene, WSe$_2$, graphene and hBN flakes from their substrate (oxidized silicon). hBN flakes are typically 15 to 50 nm thick, depending on the device. The resulting stack is deposited onto a degenerately doped silicon substrate covered with a 285 nm thick SiO$_2$ layer, which we use as a back gate ($V_G$). Top and bottom graphene flakes are electrically connected by one-dimensional contacts[33] made of 2 nm Ti/ 100 nm Au.

**Optoelectronic measurements.** Photocurrent is generated by focusing a laser beam (close to diffraction limit) with a microscope objective (Olympus LUCPlanFLN 40x) on the device, and measured with a preamplifier and a lock-in amplifier synchronized with a mechanical chopper. A supercontinuum laser (NKT Photonics SuperK extreme) with tunable wavelength (from 500 to 1500 nm), pulse duration of ~40 ps and repetition rate of 40 MHz is employed to characterize the devices. Time-resolved photocurrent measurements are performed with ~200 fs pulses (at the sample, not Fourier-transform-limited) centered at 800 nm (1.55 eV) with a spectral bandwidth of ~200 nm. Since the laser spectrum overlaps with the A exciton peak (Figure 1e), we expect a significant contribution from these excitons to the photocurrent generated with this laser. Laser pulses are generated by a Ti:Sapphire laser (Thorlabs Octavius) with a repetition rate of 85 MHz. The optical beam is split into two arms and recombined using 50/50 beamsplitters. One arm contains a motorized translation stage that allows for the generation of a computer-controlled time delay *Δt* between the two pulses.

**Extraction of the photoresponse time.** Global fitting of the time-resolved photocurrent signals is done using the following equation:

$$\frac{PC(\Delta t)}{PC(\Delta t \to \infty)} = 1 - A \exp\left(\frac{-|\Delta t|}{\tau}\right) + B \exp\left(\frac{-|\Delta t|}{\tau_G}\right)$$

where amplitudes *A* and *B*, and time constants $\tau$ and $\tau_G$ are the fitting parameters. The exponential with time constant $\tau_G$ is an empirical term introduced to reproduce the flattening of the signals at low time delay ($|\Delta t| \leq 2.2$ ps) which is observed on the thinnest devices ($L$ < 5 nm). This contribution is always small compared to the other exponential term ($|A|$ >> $|B|$) and slightly faster ($\tau_G$ ~ 2-5 ps). We attribute this signal to the optoelectronic response of the top and bottom graphene layers[7]. We emphasize that this small correction to the global exponential decay is introduced for the sake of generality of the fitting procedure, and in any case does not influence the extracted response time τ and subsequent conclusions.

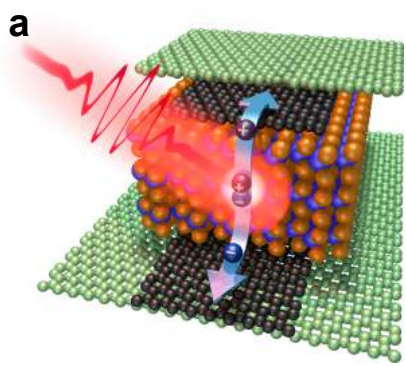
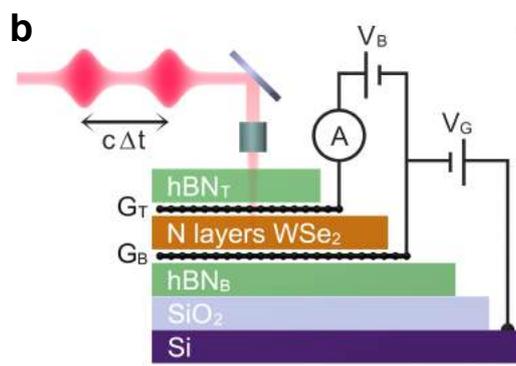
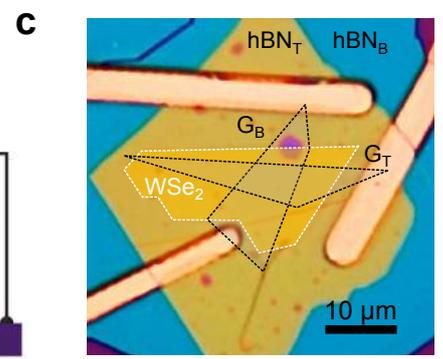
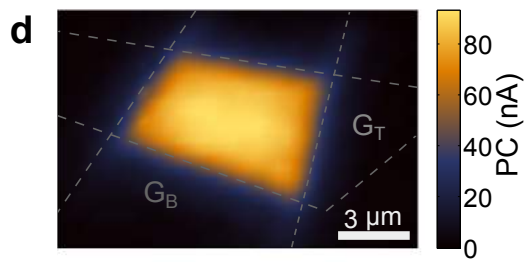
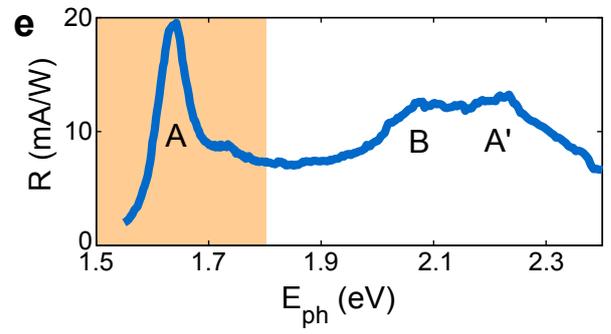

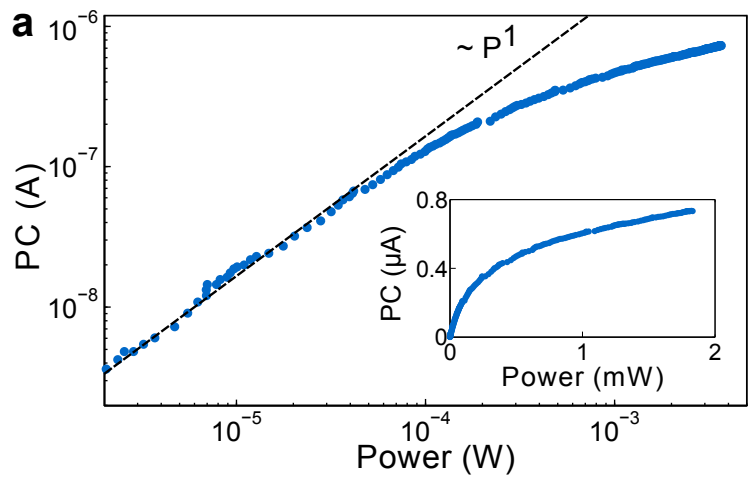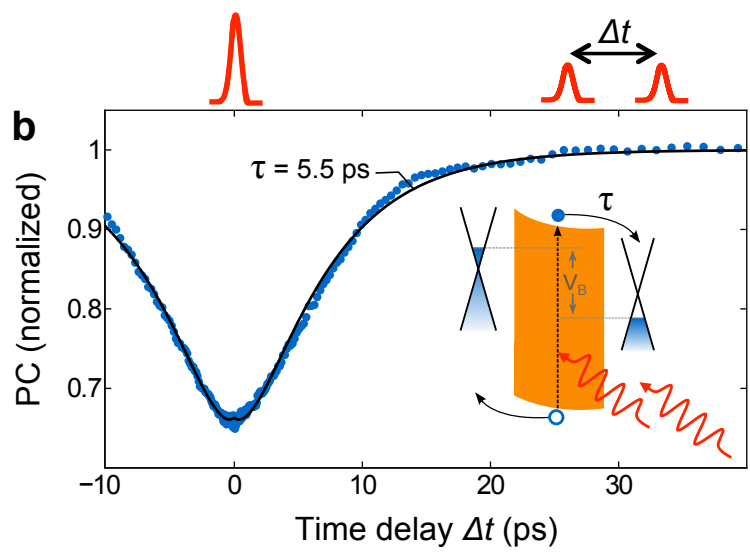

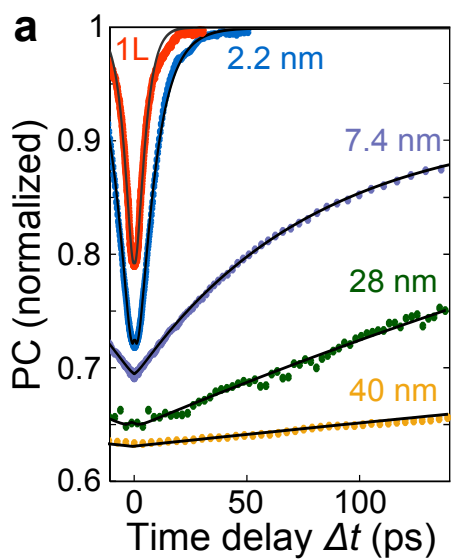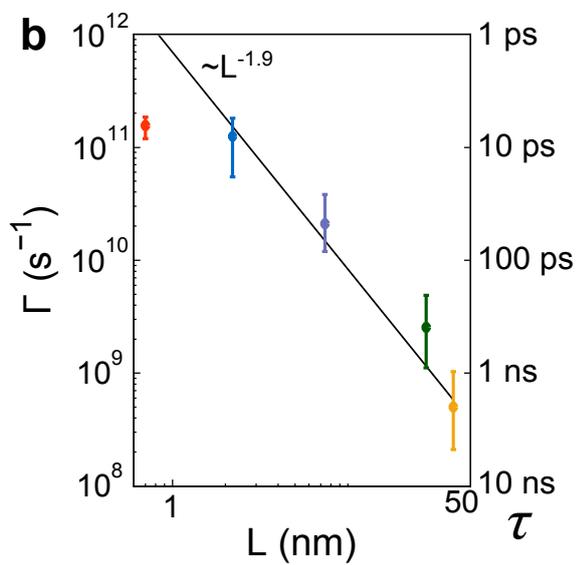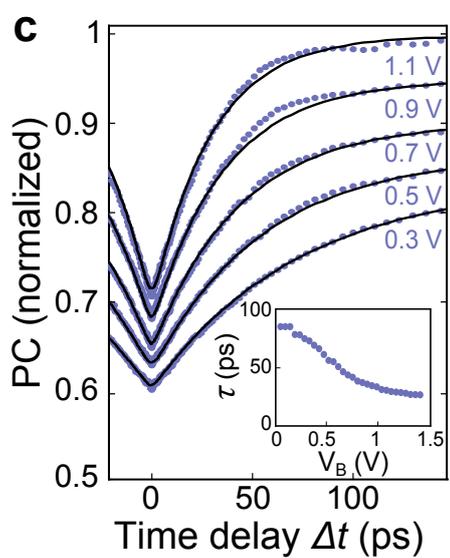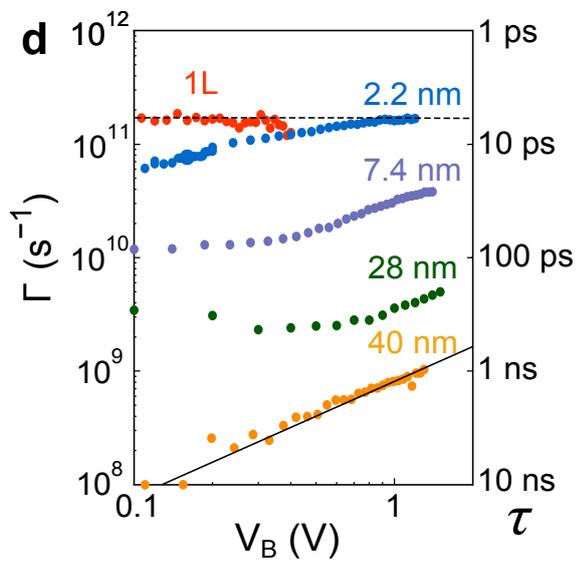

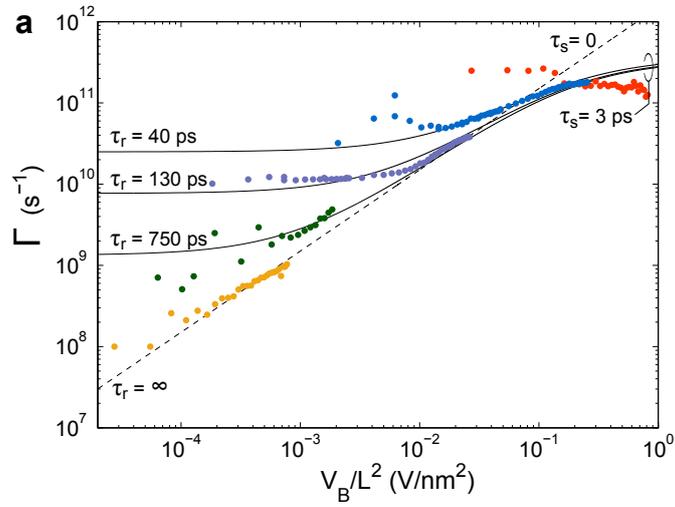
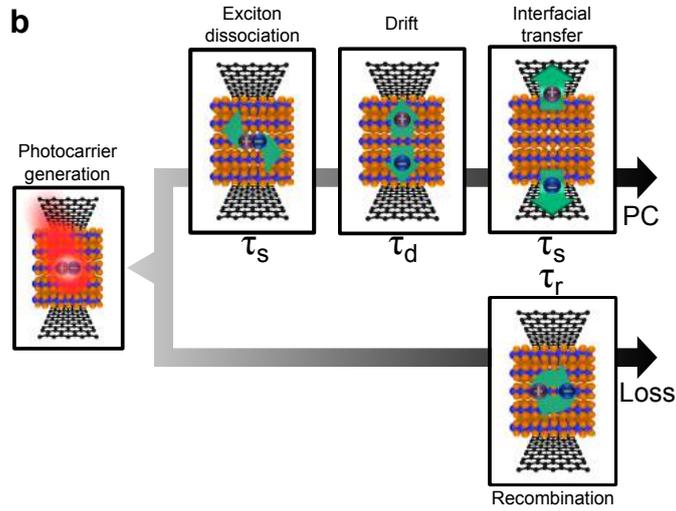
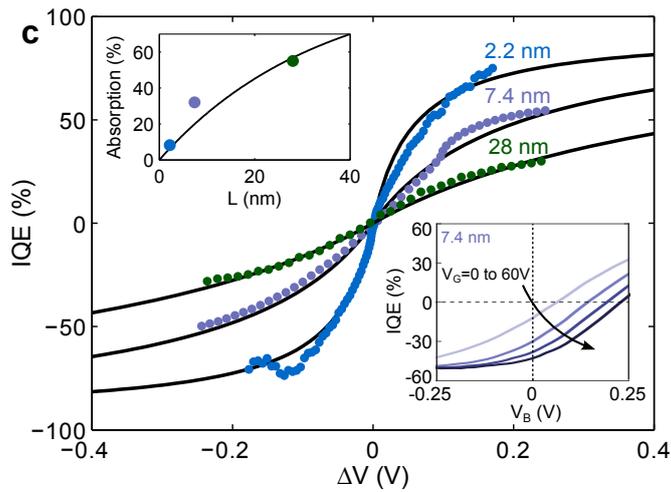

# Picosecond photoresponse in van der Waals heterostructures

M. Massicotte, P.Schmidt, F. Vialla, K. Watanabe, T. Taniguchi, K.J. Tielrooij, F.H.L. Koppens

## I  Electrostatic model of the G/WSe$_2$/G heterostructure

Here we show that the photocurrent is driven by the out-of-plane potential drop inside WSe$_2$. For this we use an electrostatic model, similar to the one used by Yu *et al.* [1], that predicts the band alignment of the graphene-WSe$_2$-graphene heterostructures.

The studied devices are deposited onto a substrate made of two different dielectrics, the thermally grown SiO$_2$ with thickness $D_{\text{SiO}_2} = 285$ nm, and a hBN flake (15 nm $< D_{\text{hBN}} <$ 50 nm). Therefore, we add the capacitances per unit area $C_{\text{G}}$, $C_{\text{SiO}_2}$ and $C_{\text{hBN}}$ of graphene, SiO$_2$ and hBN

$$\frac{1}{C_{\text{G}}} = \frac{1}{C_{\text{SiO}_2}} + \frac{1}{C_{\text{hBN}}} = \frac{D_{\text{SiO}_2}\epsilon_{\text{hBN}} + D_{\text{hBN}}\epsilon_{\text{SiO}_2}}{\epsilon_0 \epsilon_{\text{SiO}_2} \epsilon_{\text{hBN}}}. \tag{1}$$

Here $\epsilon_0$ is the vacuum permittivity and $\epsilon_{\text{SiO}_2} = 3.9$ and $\epsilon_{\text{hBN}} = 4.2$ [2] are the relative permittivities of SiO$_2$ and hBN.

Apart from the electric field induced by the backgate voltage $V_G$, there are additional fields at the graphene-WSe$_2$ interfaces, which are induced by charge transfer from WSe$_2$ to graphene. We call them $E_B$ for the bottom graphene flake and $E_T$ for the top graphene flake. Taking these fields into account, the charge densities in bottom and top graphene are given by

$$en_{\text{B}} = C_{\text{G}} V_{\text{G}} - \epsilon_{\text{WSe}_2} \epsilon_0 E_{\text{B}} \tag{2}$$

$$en_{\text{T}} = \epsilon_{\text{WSe}_2} \epsilon_0 E_{\text{T}}, \tag{3}$$

where $\epsilon_{\text{WSe}_2} = 4.2$ is the relative permittivity of WSe$_2$ [3]. The fields at the top and bottom graphene-WSe$_2$ interfaces have the following relationship

$$E_{\text{T}} = E_{\text{B}} + \frac{NeL}{\epsilon_{\text{WSe}_2}}, \tag{4}$$

where $N$ is the donor density in WSe$_2$ and $L$ is the thickness of WSe$_2$. We assume that WSe$_2$ is fully depleted. From the electric fields we calculate the electric potential and a potential drop $\Delta V$ inside WSe$_2$:

$$\Delta V = \frac{1}{2}(E_{\text{B}} + E_{\text{T}})L. \tag{5}$$

Finally, we calculate the chemical potential in each of the two graphene flakes as

$$\mu = \hbar v_{\text{F}} \sqrt{\pi |n + n_0|}, \tag{6}$$

where n$_0$ is the intrinsic charge density of the graphene flakes. When a bias voltage V$_{\text{B}}$ is applied, the sum of the difference in Fermi levels between the top and bottom graphene layers and the potential drop inside WSe$_2$ has to be equal to the applied bias voltage:

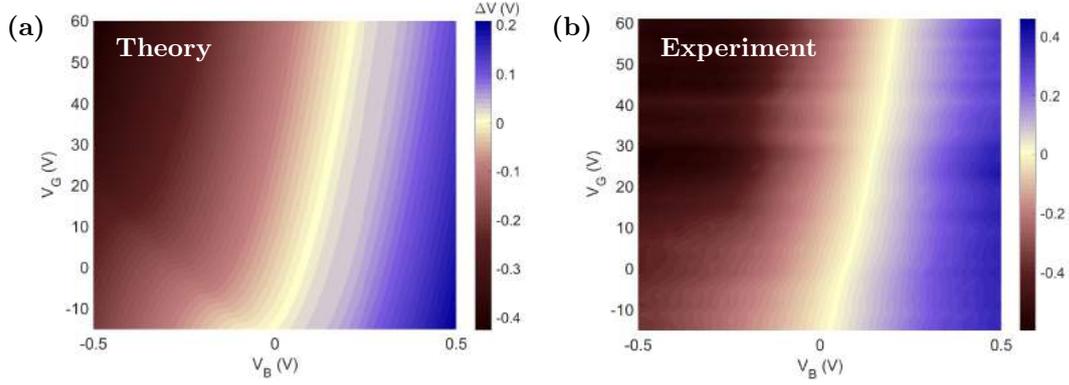

Supplementary Fig 1: **(a)** Potential drop $\Delta V$ in WSe$_2$, calculated from the electrostatic model presented in the text. **(b)** Photocurrent as a function of bias and backgate voltages for the device with a WSe$_2$ flake thickness $L = 7.4$ nm. The data was taken at a temperature of 26 K, an illumination wavelength of 575 nm and an illumination power of 10 $\mu$W.

$$eV_\text{B} = e\Delta V + \mu_\text{T} - \mu_\text{B}. \qquad (7)$$

The resulting potential drop $\Delta V$ is plotted in Supplementary Figure 1a. Its bias and gate voltage dependences follow very well that of the measured photocurrent (Supplementary Figure 1b), with the only fitting parameters being the donor density of WSe$_2$ ($N = 8\text{x}10^{17}$ cm$^{-3}$) and the initial doping of the two graphene flakes ($n_{0,B} = 8\text{x}10^{11}$ cm$^{-2}$ for the bottom graphene and $n_{0,T} = 0$ for the top graphene for the device with a WSe$_2$ flake thickness of $L = 7.4$ nm - these values vary slightly for each device). From the measured and simulated data we deduce that the photocurrent is proportional to the potential drop inside WSe$_2$ and that this potential drop can be controlled by $V_\text{G}$ and $V_\text{B}$ [1,4].

We now look closer at the short-circuit current $I_\text{SC}$ and open-circuit voltage $V_\text{OC}$. $I_\text{SC}$ is defined as the current at $V_\text{B} = 0$ and we assume it to be proportional to $\Delta V$, $I_\text{SC} = \alpha \Delta V$, with the proportionality constant $\alpha = 2$ M$\Omega^{-1}$ used as a fitting parameter. The open-circuit voltage $V_\text{OC}$ is defined as the bias voltage at which the photocurrent vanishes. This happens when there is no net electric built-in potential inside WSe$_2$. We find an excellent agreement between theory and experiment (Supplementary Figures 2a and 2b), further confirming that the potential drop $\Delta V$ inside WSe$_2$ drives the photocurrent.

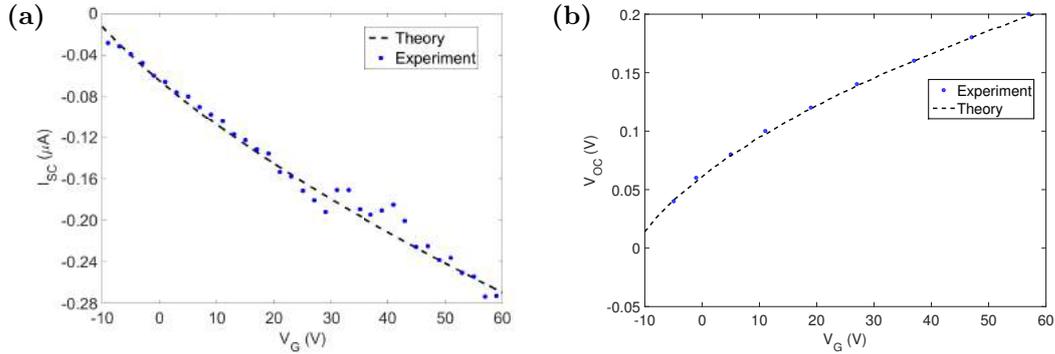

Supplementary Fig 2: **(a)** Measured short-circuit current $I_\text{SC}$ and fit to theory $I_\text{SC} = \alpha \Delta V$, where $\alpha = 2$ M$\Omega^{-1}$ is a fitting parameter. **(b)** Measured open-circuit voltage $V_\text{OC}$ and simulated values for the bias voltage at which the electric field inside WSe$_2$ vanishes. The data shown were taken at the same experimental conditions as described in Figure 1b.

## II  Power dependence of photocurrent

We perform measurements of the power dependence for every G/WSe$_2$/G device. For this we sweep the excitation power over 4 orders of magnitude at different excitation wavelengths using a supercontinuum laser (superK). Supplementary Figures 3a and 3b present the results for devices with WSe$_2$ thicknesses of $L = 2.2$ and 7.4 nm. We observe similar behaviors with clear sublinearity at high excitation power. We find an identical trend with power when excitation is done with the pulsed Ti:Sapphire laser used in the time resolved measurements of the main text.

Remarkably, if we consider each device separately, we find an identical power dependence for the different probed excitation wavelengths when the excitation power is normalized by a wavelength dependent factor that stands for the different absorption coefficients. This is evidenced by the horizontally translated curves in the log-log scale and the unique photocurrent value where the non-linearity begins. If we compare the different devices with excitation at the A exciton resonance (Supplementary Figures 3c and 3d), we find that the nonlinear regime starts for slightly different powers, with a global trend of a less marked nonlinearity the thinner the WSe$_2$ flake. Interestingly, we get very similar saturation thresholds when the excitation power is normalized by the absorption values extracted from the final analysis in the main text (Figure 4c inset). These observations are compatible with the different mechanisms mentioned in the main text and do not allow to discriminate between each one of them. However, we note that the actual origin of the sublinearity does not influence our claim that the time-resolved photocurrent measurements directly reflect the dynamics of photocarriers in the system.

## III  Real-time photoresponse and RC-time constant

We directly measure the photo-switching rate of our photodetectors by probing their response to a short light pulse in real time. Using a fast oscilloscope (Agilent MSO9404A, 4 GHz), we monitor the voltage across a 50 $\Omega$ load resistance ($R_L$ in Supplementary Figure 4d). We find a similar bias-independent response time of 1.6 ns ($1/\tau \sim 0.6$ GHz) for both the 2.2 and 7.4 nm devices (Supplementary Figures 4a to c). These experimental values, which are limited by our electrical setup, give us a lower bound for the actual photo-switching rate of our devices, which we expect to lie in the GHz regime.

We can indeed evaluate the actual photo-switching rate of our detector. To this end, in addition to the photoresponse time discussed in the main paper, we have to take into account the resistance-capacitance (RC) time constant of the device equivalent circuit (Supplementary Figure 4d). In this vertical geometry, we expect that the RC time constant is largely determined by the large capacitance of the Gr/WSe2/Gr heterostructure, $C_{WSe_2} = \epsilon_0 \epsilon_{WSe_2} A/L$. The resistance to take into account corresponds to the two graphene sheets and the following contacts and circuits ($R_S$ in Supplementary Figure 4d). The high resistance of the WSe$_2$ channel $R_{WSe_2}$ acts here as a shunt resistance so it does not affect the RC time constant. We estimate $R_S = 1$ k$\Omega$ in our devices, yet this value can be further reduced by optimizing the circuit design, contact resistance and graphene charge density. These considerations give RC rates in the Ghz regime, which is consistent with the instrument limited photo-switching rates observed experimentally and discussed previously (Supplementary Figure 4e). We finally note that for the reasonably optimized device we propose (smaller active area $A = 5$ $\mu$m$^2$ and lower graphene and contact resistance $R = 100$ Ohm), we find much higher photo-switching rates which are RC limited only for thicknesses $L$ lower than 2 nm.

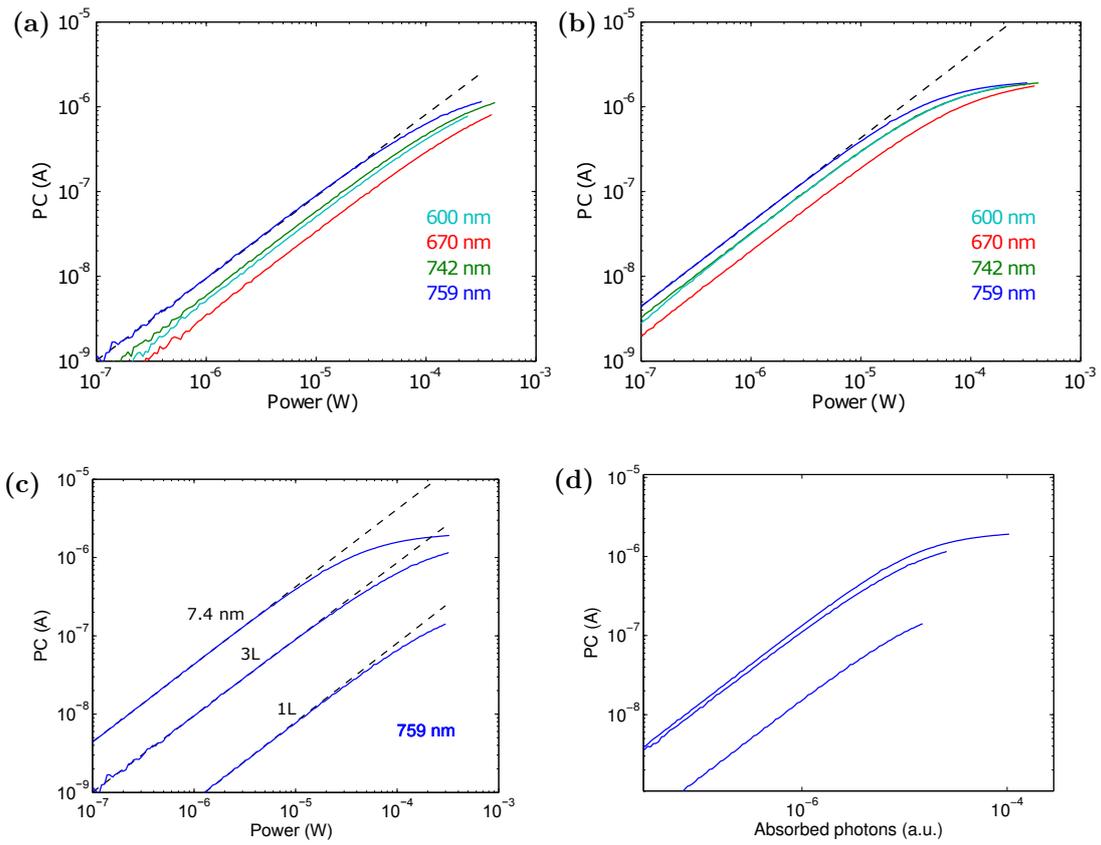

Supplementary Fig 3: Photocurrent vs optical excitation power, for devices with a (a) 2.2 nm and (b) 7.4 nm thick WSe$_2$ layer. For each device, excitation with different wavelengths gives identical trends considering the different absorptions. (c) Photocurrent vs optical excitation power for different WSe$_2$ thicknesses at a wavelength $\lambda = 759$ nm. (d) Same data as in (c) with the excitation power normalized by the absorption of the different WSe$_2$ flake thicknesses.

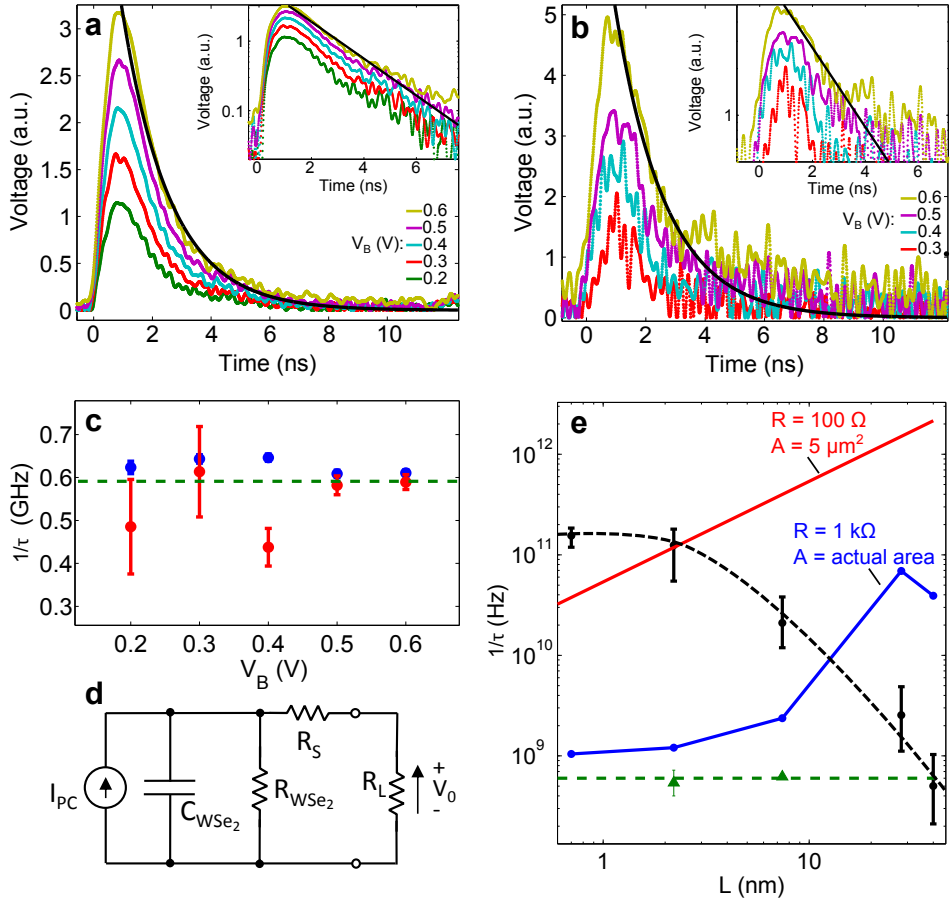

Supplementary Fig 4: **(a,b)** Real-time traces of the photoresponse to a short optical pulse (55 ps pulse duration , $\lambda = 532$ nm) for different bias voltage $V_B$ and devices with 7.4 **(a)** and 2.2 nm **(b)** thick WSe$_2$. The solid black lines are fits to the data corresponding to decay rates $1/\tau = 610$ and 590 GHz in a and b, respectively. Insets show the same data in semilog scale. **(c)** Extracted decay rates of both devices (red for 2.2 nm and blue for 7.4 nm thick WSe$_2$) as function of bias voltage. **(d)** Schematics of the equivalent circuit of our devices. **(e)** Thickness dependence of the photoresponse rate (black, the data points are measured by time-resolved photocurrent – from Figure 3b in the main text – and the dotted line is a guide to the eye), the RC time constant (evaluated for our actual devices in blue and for an optimized device in red) and the measured decay time limited by the probing setup (green, also shown in **c**).

# IV Dependence of the photoresponse time on the potential drop in WSe2

Here we show that the results we obtained from the photoresponse time model described in the main text are not affected if we consider the effective potential drop $\Delta V$ inside WSe$_2$ (Supplementary Figure 5) in our analysis instead of $V_B$ (see main text for comparison). In a first approximation $\Delta V$, which we derive from the previously presented electrostatic analysis (section I), is comparable to the applied electrical bias voltage $V_B$. We clearly observe the same trend, with a global linear dependence, and the two saturations of the response rate $1/\tau$ at high and low $V_B/L^2$ or $\Delta V/L^2$. Moreover, the time constants ($\tau_s$ and $\tau_r$) and mobility ($\mu$) values we extract in Supplementary Figure 5 are very similar to those reported in the main text. Therefore, the different discussions and conclusions of the main text hold true even when these small electrostatic deviations are taken into account.

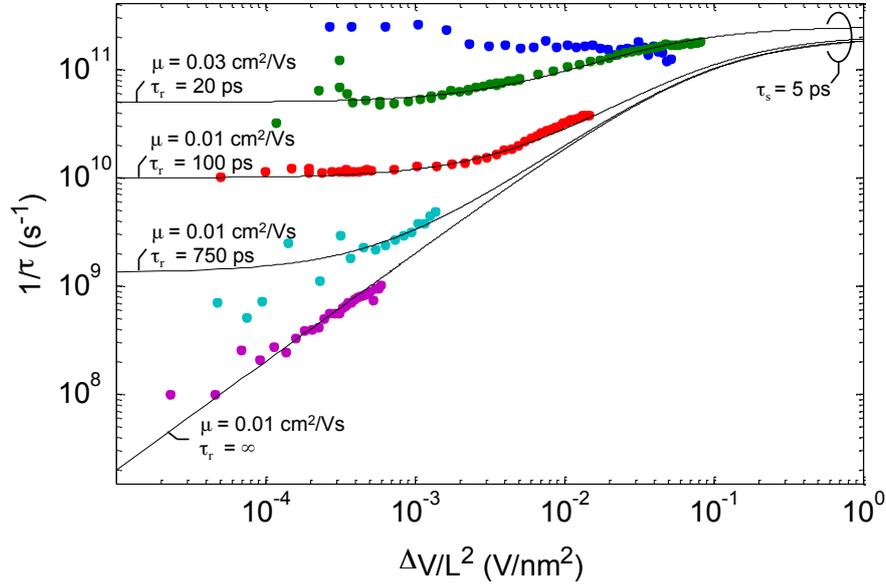

Supplementary Fig 5: Response rate $1/\tau$ vs $\Delta V/L^2$. The potential drop in WSe$_2$ ($\Delta V$) was calculated using the electrostatic model described in the text. The solid black lines correspond to the time response model $\tau^{-1} = (L^2/2\mu\Delta V + \tau_s)^{-1} + \tau_r^{-1}$. The fitted values are indicated on the figure for each curve.

# V Photoluminescence quenching

We perform spatial mapping of the photoluminescence from WSe$_2$ on the 1L and 3L devices on a different experimental setup than the one for photocurrent measurements. We use 532 nm excitation light (diffraction limited excitation spot with power $\simeq 50$ $\mu$W) and a detection with either a single photon counter module (coupled with longpass filters) or a spectrometer (coupled to an EMCCD camera), in a confocal geometry. Spatial maps of the photoluminescence intensity are presented in Supplementary Figure 6, as well as corresponding photoluminescence spectra at given laser positions. Relative intensity and spectral lineshape from the two devices confirm the respective 1L and few-layers WSe$_2$ geometry [5].

We clearly observe a strong reduction of the luminescence from the areas where WSe$_2$ is in contact with the top and/or bottom graphene layers. This emission quenching is a strong evidence for an energy transfer from the WSe$_2$ to the graphene, by the addition of a highly efficient non-radiative relaxation pathway for the photogenerated excitons [6].
A quantitative evaluation of the quenching ratios shows on average a 300-fold and 10-fold
reduction of the luminescence in the 1L and 3L device respectively. This strong thickness
dependence is indicative of an energy transfer mechanism. For instance in a Förster resonant
energy transfer formalism, where dipole-dipole interaction between a point-like emitter and
a plane is evaluated, the induced decay time of the emitter scales with the distance to the
power of 4 [6, 7]. Moreover, the photoluminescence intensity being driven by the lifetime of
the excitons in the $WSe_2$, the quenching ratio directly gives the ratio between the intrinsic
lifetime (without graphene) and the reduced lifetime due to the energy transfer to graphene.
Exciton lifetimes in $WSe_2$ have already been evaluated to be hundreds of ps from several
optical pump-probe studies [8]. This gives reduced lifetimes of roughly tens of ps for the 3L
device and a few ps for the 1L device. This timescale should correspond to the zero-bias
photoresponse time of the device, introduced as $\tau_r$ in the main text. It appears to be in good
agreement with the values extracted from the time-resolved photocurrent measurements.

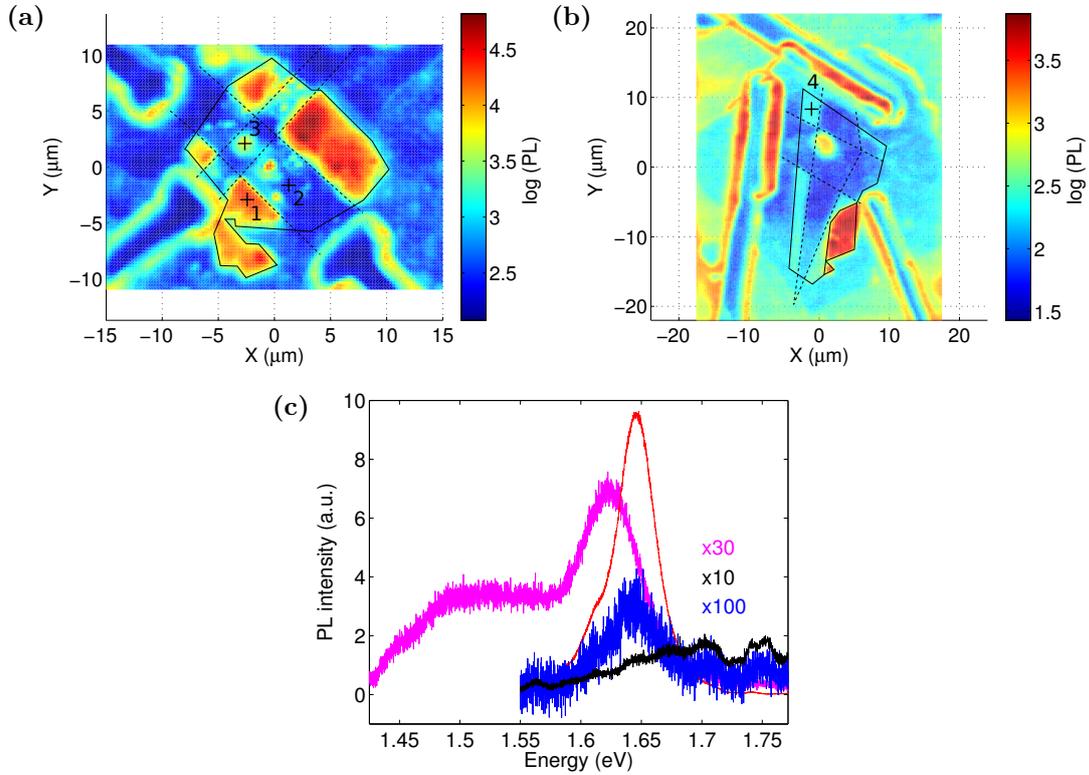

Supplementary Fig 6: $WSe_2$ photoluminescence intensity (in log scale) spatial maps measured
on the devices with a $WSe_2$ flake (drawn with solid black lines) consisting of **(a)** a monolayer
and **(b)** 3 layers. Strong quenching is observed in the area in contact with the graphene flakes
(drawn with dashed lines). **(c)** Photoluminescence spectra obtained on the marked spots in
the maps, showing strong and spectrally narrow luminescence from the monolayer (1, red),
quenched luminescence from the same monolayer (2, blue), scattering or luminescent defects
(3, black) and spectrally broad luminescence from the 3-layer flake (4, pink). We note the
inclusion of a strongly luminescent monolayer flake on the bottom edge of the 3-layer flake.

# VI Characterization of the photodetectors

In this section we present an overall characterization of our devices as photodetectors by presenting their responsivity ($\Re$), external quantum efficiency (EQE), noise equivalent power (NEP) and specific detectivity (D*) in Supplementary Table 1.

We calculate the EQE (the number of electrons contributing to the photocurrent per incident photon) as EQE = $\frac{PC}{P}\frac{hc}{e\lambda}$ with $PC$ being the measured photocurrent, $P$ the illumination power at the sample, $\lambda$ the illumination wavelength and $h$, $c$ and $e$ the Planck constant, speed of light in vacuum and electron charge. The EQE of the device with a WSe$_2$ flake thickness of $L = 7.4$ nm is presented as function of the bias voltage $V_B$ in Supplementary Figure 7. In Supplementary Table 1, we note that higher values for responsivity and EQE can be reached for the device with $L = 40$ nm when applying very high bias and backgate voltages. However, we did not apply these high voltages in order to protect the device.

| L (nm) | $\Re$ (mA/W) | EQE (%) | NEP (W/Hz$^{1/2}$) | D* (cm Hz$^{1/2}$/W) |
|---|---|---|---|---|
| Monolayer | 1.8 | 0.29 | 1x10$^{-10}$ | 4x10$^6$ |
| 2.2 | 44 | 7.3 | 5x10$^{-12}$ | 1x10$^8$ |
| 7.4 | 110 | 18 | 1x10$^{-12}$ | 1x10$^9$ |
| 28 | 110 | 18 | 6x10$^{-13}$ | 6x10$^8$ |
| 40 | >80 | >13 | <1x10$^{-14}$ | >5x10$^{10}$ |

Supplementary Table 1: Figures of merit for the different devices. Light excitation wavelength and power are 759 nm and 1 $\mu$W, respectively. Values of NEP and D* are evaluated at $V_B = 0$ and $V_G = 30$ V.

We can optimize the NEP by taking advantage of the electrical tunability of our devices with bias $V_B$ and gate voltage $V_G$. As seen in the previous section, applying a gate voltage $V_G$ increases the open circuit voltage. It is therefore possible to reach high photocurrent efficiency without applying any bias $V_B$. In this configuration, we minimize the dark current $I_d$ and the noise of the detector is no longer dominated by the shot noise but only by the thermal contribution (Supplementary Figures 8ab). We evaluate the NEP as NEP = $\frac{1}{\Re}\sqrt{2eI_d + \frac{4k_BT}{R_d}}$, where $k_B$ is the boltzmann constant, $T$ the temperature and $R_d$ the equivalent resistance extracted from the dark current I-V curve at the operating bias [9]. For the device with $L = 7.4$ nm, we find values lower than $10^{-11}$ W/Hz$^{1/2}$ even at high bias where the highest efficiency is reached (Supplementary Figure 8c). Using a gate voltage $V_G = 30$ V, the NEP goes down to $10^{-12}$ W/Hz$^{1/2}$ at zero bias voltage while the IQE remains high ($\sim$40%). These values of NEP are similar to those of typical commercial photodetectors.

The specific detectivity is defined as D* = $\frac{\sqrt{A}}{NEP}$ where $A$ is the photoactive area of the device and is also reported in Supplementary Table 1.

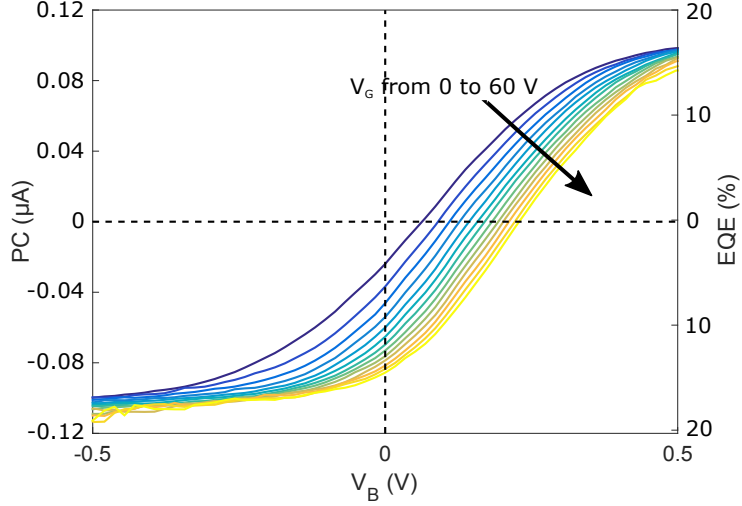

Supplementary Fig 7: Photocurrent $PC$ as a function of $V_B$ for different $V_G$. Shown are the measured values for the device with a $WSe_2$ thickness $L = 7.4$ nm. The data were taken at room temperature with an illumination wavelength $\lambda = 759$ nm and illumination power $P = 1$ $\mu$W. Also indicated on the right axis is the external quantum efficiency (EQE) corresponding to the photocurrent.

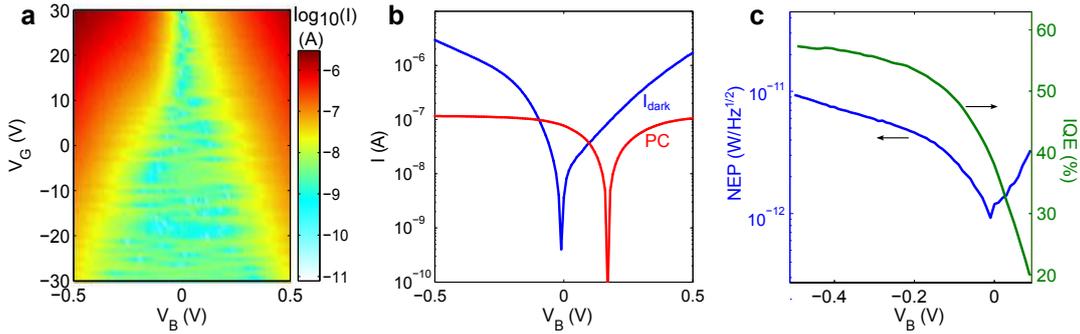

Supplementary Fig 8: **(a)** Bias and gate voltage dependence of the dark current for the $L = 7.4$ nm device. **(b)** Dark current (blue) and photocurrent (red) as function of bias voltage at gate voltage $V_G = 30$ V. Excitation wavelength and power are 759 nm and 1 $\mu$W, respectively. **(c)** Noise equivalent power (blue) and internal quantum efficiency (green) as function of bias voltage (same conditions as **b**).

# VII  Dependence of photocarrier mobility on laser power

For sake of completeness, we study the effect of laser power on the extracted photoresponse time $\tau$ of the devices and the underlying photocarrier dynamic processes. As a general rule, we observe that higher power leads to slower photoresponse time. Supplementary Figure 9 shows this effect for 4 different laser powers measured on the device with WSe$_2$ layer thickness $L = 7.4$ nm.

Using the photoresponse time model described in the main text, we find that laser power mostly affects the mobility $\mu$ of the photocarriers, which varies from 0.002 to 0.017 cm$^2$/Vs as the average laser pulse power decreases from 1000 to 55 $\mu$W. The reduction of the mobility with increasing power (and therefore higher photocarriers densities) suggests that photocarriers mobility is limited by carrier-carrier scattering. This interpretation is consistent with the assumption that carrier-carrier interactions are responsible for the sublinear power dependence. Since the photocurrent autocorrelation technique relies on this sublinear effect, a sufficiently high laser power is required in order to reach the sublinear regime and extract a reliable response time. Hence, the values of mobility (response time) reported in the main text should be regarded as lower (upper) bounds.

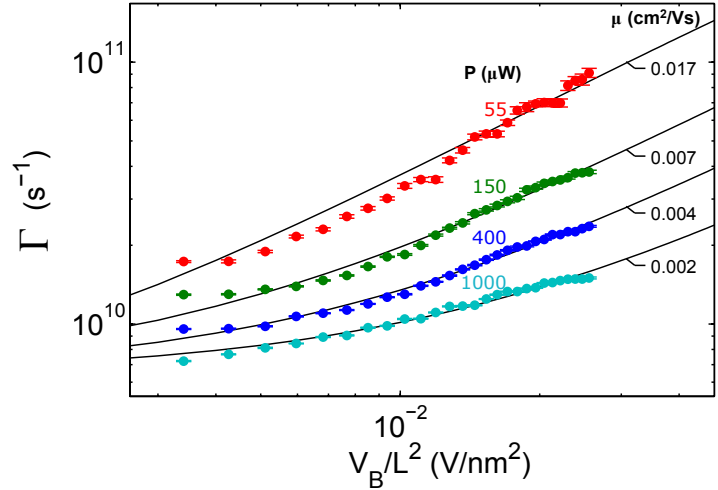

Supplementary Fig 9: Photoresponse rate ($\Gamma = 1/\tau$) vs $V_B/L^2$ obtained from the device with a WSe$_2$ layer thickness $L = 7.4$ nm, for different average laser pulse power $P$ (indicated on the figure). The solid black lines correspond to the time response model $\tau^{-1} = (L^2/2\mu V_B + \tau_s)^{-1} + \tau_r^{-1}$, with $\tau_s = 1 - 3$ ps and $\tau_r = 130 - 200$ ps. The fitted mobility $\mu$ is indicated for each curve.